\documentclass[11pt]{article}
\usepackage{moriond,epsfig}

\def\Journal#1#2#3#4{{#1} {\bf #2}, #3 (#4)}

\def\mco{\multicolumn}

\def\be{\begin{equation}}
\def\ee{\end{equation}}
\def\bea{\begin{eqnarray}}
\def\eea{\end{eqnarray}}

\def\epem{e^+e^-}
\def\mumu{\mu^+\mu^-}
\def\jpsi{J/\psi}
\def\jpsipp{J/\psi\pi^+\pi^-}
\def\jpsiw{J/\psi\omega}
\def\psip{\psi^\prime}
\def\etac{\eta_{_c}}
\def\chizero{\chi_{_{c0}}}
\def\chione{\chi_{_{c1}}}
\def\chitwo{\chi_{_{c2}}}
\def\to{\rightarrow}

\def\ccbar{c\bar{c}}

\def\DDbar{D\bar{D}}
\def\DDpi{D^0\bar{D^0}\pi^0}

\def\BBbar{B\bar{B}}

\def\jpc{J^{PC}}
\def\Kshort{K^0_S}

\def\to{\rightarrow}

\begin{document}
\vspace*{4cm}
\title{NEW CHARMONIUM-LIKE STATES AT BABAR AND BELLE}

\author{R.Mussa}

\address{INFN Sezione di Torino, Via Giuria 1 \\
I-10125 Torino, Italy}

\maketitle\abstracts{
BaBar and Belle have recently revived the interest in charmonium spectroscopy, 
discovering many unexpected resonances. In this review, I'll focus mostly 
 on the states found in B decays and double $\ccbar$. A better understanding of 
their production mechanism can help to discriminate among models, confirm
tentative $J^{PC}$ assignments, and clarify the overall picture. 
}
\section{Introduction}

Since the advent of asymmetric B factories, charmonium spectroscopy is 
living a second renaissance. Beside  the discovery of long sought \cite{QWG-YR:2005} 
$\eta_c$(2S) and $h_c$(1P), more than half a dozen new states were observed,
above the open flavor thresholds. This review will only cover the new
charmonium-like resonances observed in B decays, in $\gamma\gamma$, and 
in $\epem\rightarrow\ccbar\ccbar$. The new vector charmonia found in ISR will 
be covered by a separate talk \cite{Grauges}.

\section{Charmonium and charmonium-like objects in B decays}

The $B\to(\ccbar)K$ decays have been 
extensively studied to investigate the CP violation in weak interactions;
as first 'byproduct', came the discovery of the 'true' $\eta_c(2S)$ state.
Studies of the multibody processes with a $J/\psi$, a kaon and 
other light hadrons led to the discovery of three other resonances, 
named with the last letters of the alphabet: X(3872), Y(3940), Z(4430).
Their nature is still unclear, and their properties are 
described in the next sections.  

\subsection{X(3872): molecule, cusp or tetraquark?}

In August 2003, Belle \cite{Choi:2003ue} reported the discovery of 
 X(3872) in $B \to K X(3872) \to K \jpsi\pi^+\pi^-$.
In rapid sequence, CDF \cite{Acosta:2003zx}, D0 \cite{Abazov:2004kp}
(in  $\bar{p} p$  annihilations at $\sqrt{s}$=2 TeV)  
and BaBar \cite{Aubert:2004ns} confirmed the discovery. 
BELLE's upper limit  at 90\%CL on the total width is
 $\Gamma < 2.7 MeV$. The  PDG2006~\cite{PDG:2006} value for the X(3872) mass 
(3871.2$\pm$0.5) is  very close to  the $D^0\bar{D}^{*0}$ threshold at
3871.8$\pm$0.4 Mev/c$^2$, updated after the new measurement of the D$^0$ mass
\cite{Cawlfield:2007dw}.
The quantum numbers of X(3872) are not yet determined: 
Belle \cite{Abe:2005ix} and  BaBar \cite{Aubert:2006aj}
observed $X(3872)\to\gamma\jpsi$ (which implies C=+1)
with  a significance of 4 and 3.4 $\sigma$'s, respectively.
Both CDF \cite{Kreps:2006} and Belle \cite{Abe:2005iy} performed an angular 
analysis of $\jpsi\pi^+\pi^-$: the most likely assignments 
are $\jpc=1^{++},2^{-+}$.
Belle \cite{Abe:2005ix} claims also the observation of a 4$\sigma$ signal
in $B \to K \jpsi\pi^+\pi^-\pi^0$ with a rate comparable to the 
$\jpsi\pi^+\pi^-$ mode. The two- and three- pion mass distributions 
are clustering on the high end of the spectrum, as if produced from 
subthreshold $\jpsi\rho$ and $\jpsi\omega$ decays. 
These evidences suggest that the decay is not conserving isospin, 
or that X(3872) is not an isospin eigenstate. 
Last but not least,
Belle \cite{Gokhroo:2006xxx} and BaBar \cite{Aubert:2007xxx} 
reported evidence of decays to 
$D\bar{D}\pi$, $\approx 10\times{\cal B}(\jpsi\pi^+\pi^-)$, and 
$D\bar{D}\gamma$, $\approx 6\times{\cal B}(\jpsi\pi^+\pi^-)$.
As the $D\bar{D}\pi$ peak is 3 MeV higher than the $\jpsi\pi^+\pi^-$,
 theory speculations about a possible doublet of states have started.
On the other side, if there is only one state, the sum of 
observed decays is about to saturate the upper limit set by Babar
\cite{Aubert:2005vi} on ${\cal B}$(B $\to$ K X(3872)), obtained by searching
for monochromatic kaon recoils in a tagged B meson sample, where one 
of the two B mesons is fully reconstructed.

\subsection{Z$^\pm$(4430): the first charged resonance 
with hidden charm content}

Last summer, the Belle collaboration \cite{Belle:2007zz}
reported about one structure in the 
$B\to K \psip\pi^\pm $ Dalitz plot from a sample of 
657M $\BBbar$ pairs. Outside the known bands corresponding to 
$K^*(890)$ and $K_2^*(1430)$, a 7 $\sigma$ bump
in the $\psip \pi^\pm$ mass distribution is seen. 
The state, dubbed Z$^\pm$(4430), has a mass  M = 4433$\pm$4$\pm$1 MeV/c$^2$
and a total width $\Gamma$ = 44$^{+17}_{-13}(stat)^{+30}_{-11}(syst)$ MeV.

Four decay modes of the  $\psip$ are detected:
 $\epem$, $\mumu$ and  $\jpsi\pi\pi$ with $\jpsi\to\mumu,\epem$.
The resonance is seen both in charged and neutral B decays, 
but the significance in $B \to Z \Kshort$ does not exceed 3 $\sigma$.
If confirmed, this is the first charged tetraquark candidate, and 
the starting point of a new spectroscopy. Similar structures 
should be searched for in $\eta_c \pi K$, $J/\psi \pi K$, $\chi_c \pi K$ 
Dalitz plots. Besides the clear K$^*$  signal, statistics are such that
a comprehensive study of the  complex 
structure of three-body B decays to charmonium will probably require 
samples that are not within reach of this generation of B-factories.

\subsection{Y(3940): discovery, confirmation, doubts}

The Y(3940) is a broad resonance, $\Gamma=92\pm24$MeV,
 discovered by Belle \cite{Belle:Y3940} in B decays to $K+\omega\jpsi$.
About 1\% of the $J/\psi$'s produced in B decays come from this process.
If we hypothesize to have just one state, with reasonable assumptions on 
${\cal B}(Y(3940)\to \jpsi\omega)\approx10\%$, its partial width to 
$\jpsi\omega$ would be $\Gamma(Y(3940)\to \jpsi\omega) \approx 9 MeV $, 
much larger than the typical widths for hadronic transitions between 
chatmonia, {\it e.g.} $\Gamma(\psi^\prime\to \jpsi\pi\pi)$ =  0.16 MeV.
Large partial widths of hadronic traditions to low lying states are also
observed in the recently found vector states. 
 
The transition with emission of an $\omega$ is  unique 
in the charmonium energy range, but has  been observed by 
CLEO \cite{Severini:2004} in the bottomonium system: 
$\chi_{_{b1,2}}(2P)\to\Upsilon(1S)$. 
 
Recently, BaBar \cite{Mokhtar}
has confirmed the observation of a peak in $\jpsi\omega$,
but  narrower and at a lower energy. The analysis is based on a slightly 
larger sample, 348 fb$^{-1}$, and gives 
M=3914.6$\pm$3.6$\pm$1.9 MeV/c$^2$ and 
$\Gamma$ = 33$\pm$10$\pm$5 MeV.  
While it is simple to isolate the $\omega$ peak in the $3\pi$ system  at
higher masses, in the region below 4 GeV also the 
modeling of the phase space may induce some large systematic error.
     The $\jpsi\omega$ final state is accessible 
from almost all possible $\ccbar$ quantum numbers, and even an angular analysis
would be unreliable, if more states are merging in the same region.
The Y(3940) signal still needs to be clarified experimentally, before handing 
it over to theory speculations.

\subsection{Prospects}

The  quantitative picture  of the hadronization mechanism 
which leads from the $b\rightarrow \ccbar s$ current at quark level to the 
final state products is still incomplete.
 Only 25-45\% of the inclusive production of charmonia in B decays 
\cite{PDG:2006} (see Table \ref{Bdecs})
is explained as two body decay to known charmonium and a K or a K$^*$.

\begin{table}[h]
\caption{\it 
Branching ratios (units: 10$^{-4}$) for B decays to 
known charmonia (from PDG2006) and to new charmonium-like states.
In the rightmost column,
feed-down from known transitions 
is already subtracted from the inclusive rates. \label{Bdecs}}
\begin{center}
{
\vskip 0.1 in
\begin{tabular}{|l|cc|cc|c|} \hline
 ${\cal B}\times 10^4$ &  K$^\pm$ & K$^0$ & K$^{*\pm}$ &K$^{*0}$ & +anything\\ 
\hline
${\cal B}(\etac+{\cal K})$  & 9.1 $\pm$1.3 & 9.1 $\pm$1.9 & &  16 $\pm$7 & $< 90$\\
${\cal B}(\jpsi+{\cal K})$   &10.08$\pm$0.35& 8.72$\pm$0.33&14.1$\pm$0.8 & 13.3$\pm$0.6 & 
78$\pm$ 3\\
${\cal B}(\chizero+{\cal K})$ & 1.6$\pm$0.5  & $<5$ &$<28.6$ & $<7.7$ & \\
${\cal B}(\chione+{\cal K})$ & 5.3$\pm$0.7  & 3.9$\pm$0.4 & 3.6$\pm$0.9  & 3.2$\pm$0.6 &
31.6$\pm$2.5\\
${\cal B}(\chitwo+{\cal K})$ & $<0.29$ &  $<0.26$ & $<0.12$ & $<0.36$ & 16.5$\pm$3.1\\
${\cal B}(\etac(2S)+{\cal K})$ & 3.4$\pm$1.8 &   &  &  &  \\
${\cal B}(\psip+{\cal K})$  & 6.48$\pm$0.45 & 6.2$\pm$0.6 &6.7$\pm$1.4 & 7.2$\pm$0.8 &
30.7$\pm$2.1\\
${\cal B}(\psi(3770)+{\cal K})$ & $2.6 \pm 0.6$\cite{Aubert:2007xxx} & &&&\\
\hline
${\cal B}( X(3872)+{\cal K})\times{\cal B}_{\jpsipp}$ & $.114\pm .020$
& &&&\\
${\cal B}( X(3872)+{\cal K})\times{\cal B}_{\DDpi}$ &\mco{2}{|c|}{$1.41\pm 0.40$\cite{Gokhroo:2006xxx,Aubert:2007xxx}} &&&\\
${\cal B}( X(3872)+{\cal K})$ &\mco{2}{|c|}{$<2.5$\cite{Aubert:2005vi}} &&&\\
${\cal B}( Y(3940)+{\cal K})\times{\cal B}_{\jpsiw}$ & \mco{2}{|c|}{$0.15 \sim 0.71$\cite{Belle:Y3940,Mokhtar}}  &&&\\
${\cal B}( Z^\pm(4430)+{\cal K})\times{\cal B}_{\psip\pi}$ & &
$.41\pm .16$\cite{Belle:2007zz} &&&\\
\hline

\end{tabular}}
\end{center}
\end{table}

\section{Two photon physics}

Two photon scattering allows to produce C=+1 states of charmonium
with $J=even$. Recently, Belle \cite{Belle:twophoton}
 has completed a thorough study of the 
branching fractions of $\eta_c$(1S) and $\chi_{c0,2}$ to 4 charged prongs, 
and published upper limits for $\eta_c$(2S) to the same decay channels.
  

Above open charm threshold, 
Belle has probably discovered \cite{Uehara:2005qd} the $\chi_{c2}(2P)$, 
decaying to  $\DDbar$.
The measured signal (64$\pm$18 events, for a 5.3 
$\sigma$ significance) allows to calculate the product 
$\Gamma\times{\cal B}(\gamma\gamma)\times{\cal B}(D\bar{D}) = 
0.18\pm0.05\pm0.03$ keV.
A confirmation from BaBar and the measurement of its 
branching ratio to $D\bar{D}^*$  is needed.

\section{Double $\ccbar$ in $\epem$ annihilation}

Double charmonium production \cite{Yabsley:2007jb}  
in $\epem$ annihilation, first observed
by Belle in 2002, is still a puzzle for  theorists.
The mass distribution 
of objects recoiling against $\jpsi$ and $\psi^\prime$ showed clear peaks
belonging to $\eta_c,\chi_{c0}$ and $\eta_c(2S)$, discovered in B decays
few months before. 
No signal from the region with $M_{recoil}<M(\eta_c)$ is seen,
 in disagreement with NRQCD predictions, and the measured $(\ccbar)(\ccbar)$
cross section is more than five times bigger than the tree level 
calculation. 
 Table \ref{tab2c} summarizes the
updated experimental results vs NRQCD predictions (LO and NLO).

\begin{table}[h]

\caption{\it 
$\sigma(\epem\to V_{\ccbar} S_{\ccbar})$ (in fb) vs. NRQCD predictions  \label{tab2c}}

\begin{center}
{
\vskip 0.1 in
\begin{tabular}{|l|cc|cc|c|} \hline
& \mco{4}{|l|} 
{$V_{\ccbar}=\jpsi$,${\cal B}_{S_{\ccbar}>2 tracks}$} 
& 
 $V_{\ccbar}=\psip$,${\cal B}_{S_{\ccbar}>0 tracks}$ 
\\ 
\hline
  $S_{\ccbar}$ &    Belle      &  BaBar  & LO & NLO & Belle\\ 
\hline
$\eta_c(1S)$   &  25.6$\pm 2.8\pm 3.4$  &  17.6$\pm 2.8^{+1.5}_{-2.1}$ & 
3.78$\pm$1.26 & 17.6$^{+7.8}_{-6.3}$ &  16.3$\pm 4.6\pm 3.9$\\
$\chi_{c0}$    &  6.4$\pm 1.7\pm 1.0$  &  10.3$\pm 2.5^{+1.4}_{-1.8}$ & 
2.40$\pm$1.02 &   &              12.5$\pm 3.8\pm 3.1$      \\
$\eta_c(2S)$   &  16.5$\pm 3.0\pm 2.4$  &  16.4$\pm 3.7^{+2.4}_{-3.0}$  & 
1.57$\pm$1.52 &   &        16.0$\pm 5.1\pm 3.8$            \\
\hline
\end{tabular}
}
\end{center}

\end{table}

\noindent The  comparison  with theory at  full NLO \cite{Bodwin:2007ga}
is  possible only for the $\jpsi\eta_c$ channel: 
the resummation of  ${\cal O}(\alpha_s)$ terms contributes an extra 80\%,
and ${\cal O}(v^2)$ terms give another  145$\pm$61\%.

The $\jpsi$ recoil method was further refined by Belle, to exploit 
the dominant decay of states above threshold to D mesons.
Reconstructing a large fraction of both charged and neutral D mesons, 
and exploiting the constraint on M(D), it is possible to single out 
the D and D$^*$ peaks , and resolve
the exclusive processes $\epem\to\jpsi D^{(*)}\bar{D}^{(*)}$.
This allowed to improve the resolution on mass and width of
two newly discovered  states, named X(3940) \cite{Belle:X3940}, 
and  X(4160) \cite{Belle:X4160}, 
and to measure their branching fractions to open charm mesons
(see table \ref{tabx}).

\begin{table}[h]
\caption{\it Properties of the new states found in double 
$\ccbar$, measured by Belle from a sample of 693 fb $^{-1}$
 \label{tabx}}

\begin{center}
{
\vskip 0.1 in
\begin{tabular}{|l|c|c|c|c|c|} \hline

State & signif. & M(MeV/c$^2$) & $\Gamma$(MeV) & decay & 
$\sigma(\jpsi X)\times {\cal B}_{out}$(fb) \\ 
\hline
X(3940) & 6.3$\sigma$ &  $3942^{+7}_{-6}\pm 6$  &   
$37^{+26}_{-15}\pm 8 $ & $D\bar{D}^*$ & $13.9^{+6.4}_{-4.1}\pm 2.2 $ 
\\
X(4160) & 5.4$\sigma$ &  $4156^{+25}_{-20}\pm 16$  & 
$139^{+111}_{-61}\pm 22 $ & $D^*\bar{D}^*$ & $24.7^{+12.8}_{-8.3}\pm 5.0 $
\\ 
\hline
\end{tabular}}
\end{center}
\end{table}

Conventional charmonium interpretations for these states would point to 
the  $\eta_c(3S)$ and $\chi_{c0}(3P)$ states. 
A confirmation of these  states by BaBar is needed, and possibly an angular
analysis should be performed to make firmer assessment on the quantum numbers.

\section*{Acknowledgments}
 I'm grateful to my colleagues in Belle and BaBar, who asked me to give 
this review, and to all the friends in  CLEO, CDF and BES  who contributed 
to the heavy quarkonium renaissance. Thanks also to all theorists in the 
QWG, for many exciting discussions in the recent years.

\end{document}